\documentclass[conference]{IEEEtran}
\IEEEoverridecommandlockouts
\usepackage{cite}
\usepackage{amsmath,amssymb,amsfonts}
\usepackage{algorithmic}
\usepackage{graphicx}
\usepackage{textcomp}
\usepackage{xcolor}
\usepackage{subcaption}
\usepackage{enumitem}
\usepackage{threeparttable}
\def\BibTeX{{\rm B\kern-.05em{\sc i\kern-.025em b}\kern-.08em
    T\kern-.1667em\lower.7ex\hbox{E}\kern-.125emX}}
\begin{document}

\makeatletter
\newcommand{\linebreakand}{%
  \end{@IEEEauthorhalign}
  \hfill\mbox{}\par
  \mbox{}\hfill\begin{@IEEEauthorhalign}
}
\makeatother

\title{
Anomalous Sound Detection Using a Binary Classification Model and Class Centroids
}


\author{\IEEEauthorblockN{1\textsuperscript{st} Ibuki Kuroyanagi}
\IEEEauthorblockA{\textit{Nagoya University}\\
Nagoya, Japan \\
kuroyanagi.ibuki@g.sp.m.is.nagoya-u.ac.jp}
\and
\IEEEauthorblockN{2\textsuperscript{nd} Tomoki Hayashi}
\IEEEauthorblockA{\textit{Human Dataware Lab. Co.} \\
Nagoya, Japan \\
hayashi.tomoki@g.sp.m.is.nagoya-u.ac.jp\\
hayashi@hdwlab.co.jp}
\and
\IEEEauthorblockN{3\textsuperscript{rd} Kazuya Takeda}
\IEEEauthorblockA{\textit{Nagoya University} \\
Nagoya, Japan \\
takeda@i.nagoya-u.ac.jp}
\linebreakand 
\IEEEauthorblockN{4\textsuperscript{th} Tomoki Toda}
\IEEEauthorblockA{\textit{Nagoya University} \\
Nagoya, Japan \\
tomoki@icts.nagoya-u.ac.jp}
}

\maketitle

\begin{abstract}

An anomalous sound detection system to detect unknown anomalous sounds usually needs to be built using only normal sound data.
Moreover, it is desirable to improve the system by effectively using a small amount of anomalous sound data, which will be accumulated through the system’s operation.
As one of the methods to meet these requirements, we focus on a binary classification model that is developed by using not only normal data but also outlier data in the other domains as pseudo-anomalous sound data, which can be easily updated by using anomalous data.
In this paper, we implement a new loss function based on metric learning to learn the distance relationship from each class centroid in feature space for the binary classification model.
The proposed multi-task learning of the binary classification and the metric learning makes it possible to build the feature space where the within-class variance is minimized and the between-class variance is maximized while keeping normal and anomalous classes linearly separable.
We also investigate the effectiveness of additionally using anomalous sound data for further improving the binary classification model.
Our results showed that multi-task learning using binary classification and metric learning to consider the distance from each class centroid in the feature space is effective, and performance can be significantly improved by using even a small amount of anomalous data during training.
\end{abstract}


\maketitle

\section{Introduction}
Anomalous sound detection (ASD) is the task of identifying whether a sound emitted from a particular object is normal or anomalous.
Here, an anomalous sound is caused by an atypical event, such as an accident or the malfunction or breakdown of a machine.
The detection of anomalous sounds can also be used to improve the efficiency of maintenance work on manufacturing equipment and infrastructure and to monitor equipment installed in difficult locations for people to enter.
The use of this technology is expected to become widespread during the coming fourth industrial revolution, e.g., factory automation utilizing artificial intelligence~\cite{bayram2021real,8392817}.

It would be difficult to collect data representing every possible anomalous sound because these sounds rarely occur during the normal operation of factory equipment, and the possible types of anomalous sounds are very diverse.
Therefore, when constructing an ASD model, it is often the case that only normal data is used, or that only a small amount of anomalous data is used in addition to the normal data.
One detection method that is used when only normal data is being utilized is outlier detection, which models normal data and detects data that does not correspond to the model, categorizing it as anomalous.
Typical methods include generative modeling approaches which utilize probabilistically modeling of the distribution of normal data using Gaussian mixture models~\cite{scott2004outlier}, and one-class support vector machines~\cite{tax2004support,chung2013automatic} using acoustic features such as Mel frequency cepstrum coefficients.
As a result of advances in deep learning technology, methods based on neural networks are also gaining attention~\cite{Koizumi_DCASE2020_01}.
These methods train autoencoders (AE) or autoregressive models with recursive neural networks to reconstruct normal data, and calculate the reconstruction error for use as an anomaly score~\cite{hayashi2020conformer,malhotra2015long,hayashi2018anomalous,giri2020unsupervised}.
Although these methods can achieve a high level of performance, they use only normal data during training, so it is difficult to make effective use of anomalous data.

In contrast, binary classification approaches utilize outlier sound data in addition to normal sound data such as typical target machine operating noise~\cite{ruff2020rethinking,primus2020anomalous,primus2020dcase}.
These methods assume that anomalous data is distributed outside the normal data domain, and outlier data is distributed further outside of normal data.
Based on this assumption, a binary classifier is trained using the normal data as positive examples, and the outlier data as pseudo-negative examples, so that distance from the decision boundary can be used as an anomaly score for the data.
Therefore, unlike methods that model normal data, a small amount of anomalous data, usually obtained during the ASD system’s routine operation, is directly used for training.
It is expected that these types of binary classification-based ASD methods will continuously improve due to the long-term operation of the system as more and more anomalous data is collected.

In this paper, we propose a new method for detecting anomalous sounds based on a binary classification model using outlier data.
A new loss function is introduced to the binary classification model to learn the relative distances between each sound class’s centroids in the feature space.
By using both multi-task learning of the classification task and metric learning, we should be able to map the feature space that minimizes within-class variance and maximizes between-class variance, while allowing linear separation between classes.
We also investigate the relationship between the amount of anomalous data used for training and classification performance to clarify how the amount of anomalous data used impacts the detection of anomalous sounds in binary classification models.
We conducted our experimental evaluation using the same dataset used in the DCASE~2020 Task~2 anomaly detection task~\cite{Koizumi_DCASE2020_01}.
Our results showed that 1) multi-task learning using binary classification and metric learning to consider the distance from each class centroid in the feature space is effective, and 2) performance can be significantly improved by using even a small amount of anomalous data during training.

\begin{figure}[t!]
    \captionsetup[subfigure]{position=b}
    \centering
    \begin{subfigure}[b]{0.40\textwidth}
        \includegraphics[clip, width=7.1cm, height=2.9cm]{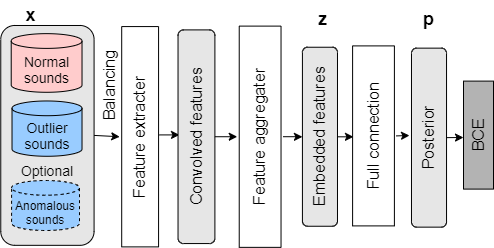}
        \caption{Classification-based method}
        \label{system2}
    \end{subfigure}
    \begin{subfigure}[b]{0.40\textwidth}
        \includegraphics[clip, width=6.0cm, height=2.9cm]{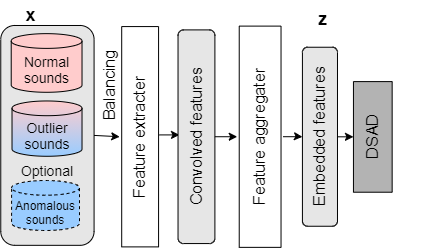}
        \caption{Distance-based method}
        \label{system1}
    \end{subfigure}
    \begin{subfigure}[b]{0.40\textwidth}
        \includegraphics[clip, width=7.1cm, height=2.9cm]{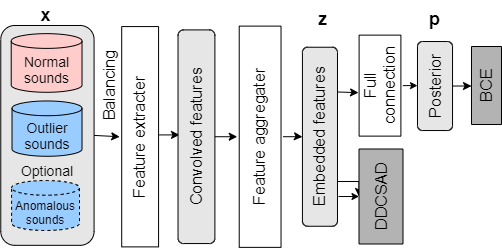}
        \caption{Proposed method}
        \label{system3}
    \end{subfigure}
    \caption{Architectures of existing and proposed ASD methods.}
    \label{three_system}
\end{figure}

\section{Related work}
This section provides a brief overview of previous research on ASD using binary classification.
We also describe a metric learning approach for improving anomaly detection, originally proposed for image processing.
\subsection{Anomalous Sound Detection Based on Binary Classification}
\label{simple_bce}
Several methods for detecting anomalous sounds based on binary classification using outlier data have been proposed~\cite{primus2020anomalous,primus2020dcase,ruff2020rethinking}.
An overview of this approach is shown in Fig.~\ref{three_system}~(\subref{system2}).
These methods assume that even task-irrelevant outlier data can be substituted as anomalous data if carefully selected.
These methods allow a model to be trained to solve a classification problem that discriminates between anomalous and normal sounds, even when anomalous data is not available.
In~\cite{primus2020anomalous}, the important aspects of outlier data selection has been identified as the matching of recording conditions, similarity to the target sound, and content diversity.

Consider a set $\mathbf{X} = \{{\bf x}_1, {\bf x}_2, ... , {\bf x}_N\}$ that has ${N}$ samples of outlier data, and a set $\tilde{\mathbf{X}} = \{\tilde{\bf x}_1, \tilde{\bf x}_2,...,\tilde{\bf x}_M\}$ that has ${M}$ samples of normal and anomalous data.
Normal and anomalous data sets are assigned labels $\tilde{y}_j \in \{+1, -1\}\ (j=1,2,...,M)$ for each data sample.
Here, $\tilde y={+1}$ indicates that the data is normal, and $\tilde y={-1}$ indicates that the data is anomalous.
The outlier data set is labeled $y \in \{+1, -1\}$, which is common to all of the data, based on a prior assumption as to whether it is closer to the normal or the anomalous data.
When performing ASD using a method based on binary classification, the network is trained to minimize the following binary cross-entropy (BCE) loss function:
\begin{equation}
    \footnotesize
    \label{eq:bce}
    \begin{split}
        &\mathcal{L}_\mathrm{BCE} = -\frac{1}{N+M}\biggl\{\sum^{N}_{i=1}{\rm log}\left(1-p_i\right) \\
        &\quad +\sum^{M}_{j=1}\{u(\tilde{y}_j){\rm log}\left(\tilde{p}_j\right)+(1-u(\tilde{y}_j)){\rm log}(1-\tilde{p}_j)\}\biggr\},
    \end{split}
\end{equation}
where $p$ and $\tilde{p}$ are the posterior probabilities output by network $\phi_p$, such as $p=\phi_{p}(\mathbf{x})$, which minimizes~\eqref{eq:bce} when $\mathbf{x}$ or $\tilde{\mathbf{x}}$ is used as input and $u(y)$ is a binary function that takes 1 for $y>0$ and 0 for $y\leq0$.
In this paper, the outlier data is always treated as pseudo-anomalous data by setting y to -1.
This assumption allows learning to be performed even when no anomalous data actually exists.
During inference, the posterior probability $p$ output by the network is used to calculate and use the anomaly score $s$ as $s=1-p$.
\subsection{Anomaly Detection Based on Metric Learning}
\label{sec_deep_sad}
Another method of anomaly detection, based on metric learning using outlier data, is deep semi-supervised anomaly detection (DSAD)~\cite{Ruff2020Deep}.
Fig.~\ref{three_system}~(\subref{system1}) shows an overview of this approach.
During classification, data that falls closer to the centroid is deemed normal, while data falling farther from the centroid are considered anomalous.
The centroid in the feature space is obtained using the pre-trained model.
The DSAD loss function is expressed as follows, using a set of $N$ outlier data $\mathbf{X}$ and a set of $M$ normal and anomalous data $\tilde{\mathbf{X}}$:
\begin{equation}
\footnotesize
\label{deep_sad}
        \mathcal{L}_{\mathrm{DSAD}} = \frac{1}{N+M}\left\{\sum^N_{i=1}\parallel\mathbf{z}_i-{\bf c}\parallel^{2y}+\eta\sum^M_{j=1}\parallel\tilde{\mathbf{z}}_j-{\bf c}\parallel^{2{\tilde{y}_j}}\right\},
\end{equation}
\noindent where $\mathbf{z}$ and $\tilde{\mathbf{z}}$ are the embedding vector output by encoder network $\phi_z$, such as $\mathbf{z}=\phi_{z}(\mathbf{x})$, which minimizes~\eqref{deep_sad} when $\mathbf{x}$ or $\tilde{\mathbf{x}}$ are used as input.
${\bf c} \in \mathbb{R}^D$ is the centroid in the feature space of the normal data $\left(\tilde{y}={+1}\cup y=+1\right)$, and $\eta > 0$ is the hyperparameter that weights the data.
For the data deemed normal, a loss is imposed on the distance between centroid $\bf c$ and the mapping point, and learning is performed to minimize the within-class variance of the data.
Note that DSAD can be used even when there is no anomalous data.
In this paper, the outlier data is always treated as pseudo-anomalous data by setting y to -1.

The training procedure is as follows.
First, the encoder network is trained as an AE using only data considered normal to obtain the parameters’ initial values.
Then, using the trained encoder, the average vector is calculated for the embedding vectors of the normal data, which is used as the centroid of the normal data.
The data deemed anomalous is also used, and learning is performed by minimizing~\eqref{deep_sad}.
Note that centroid $\bf c$ is not updated and does not change from its initial value.
During inference, the distance between embedding vector $\bf{z}$ and centroid ${\bf c}$ is used to calculate the anomaly score.

\section{Proposed Method}
\label{system_of_dcsad}
An overview of the proposed semi-supervised ASD system is shown in Fig.~\ref{three_system}~(\subref{system3}).
We propose a new loss function based on metric learning, which we call a Deep Double-Centroids Semi-supervised Anomaly Detection (DDCSAD) loss function.
The proposed loss function, which is based on metric learning, is used for the embedding vector.
The DDCSAD loss function is an extension of the DSAD loss function, that considers the centroid of normal data and the centroid of outlier data.
The DDCSAD loss function is calculated by first extending~\eqref{deep_sad} as follows:
\begin{equation}
\footnotesize
    \label{deep_dcsad}
        \begin{split}
            & \mathcal{L}_\mathrm{DDCSAD} = \frac{1}{N+M}\sum^N_{i=1}\left\{\parallel{{\bf z}_i-{\bf c}_p}\parallel^{2y}+\parallel{{\bf z}_i-{\bf c}_n}\parallel^{-2y}\right\}\\
            & \quad+\frac{\eta}{N+M}\sum^M_{j=1}\left\{\parallel{\tilde{\bf z}_j-{\bf c}_p}\parallel^{2\tilde{y}_j}+\parallel{\tilde{\bf z}_j-{\bf c}_n}\parallel^{-2\tilde{y}_j}\right\},
        \end{split}
\end{equation}
where, ${\bf c}_p\in\mathbb{R}^D$ and ${\bf c}_n\in 
\mathbb{R}^D$ represent the centroid of the normal and outlier data, respectively.
In this paper, the outlier data is always treated as pseudo-anomalous data by setting y to -1.
The following equation expresses the final loss function:
\begin{equation}
\label{eq:bce_ddcsad}
    \mathcal{L} = \mathcal{L}_\mathrm{BCE} + \lambda  \mathcal{L}_\mathrm{DDCSAD},
\end{equation}
where $\lambda > 0$ is a hyperparameter that controls the balance between the loss functions.
It is expected that multi-task learning using both the cross-entropy of the posterior probability and the DDCSAD loss function will improve the use of data and increase accuracy when learning the decision boundaries, resulting in more accurate ASD.

During training, outlier data is used as pseudo-anomaly data.
If anomalous data is available, it is also used together with the outlier data.
Unlike DSAD, the proposed method does not perform pre-training to initialize the weight parameters but instead uses randomly initialized parameters.
The initial values of the two centroids ${\bf c}_p$ and ${\bf c}_n$ are also calculated using randomly initialized parameters.
And then, they are updated at each epoch by recalculating the centroids using the entire training data set.

During inference, posterior probability $p$ (which is the output of the full connection layer), and distance ${d} = \parallel{\bf z}-{\bf c}_p\parallel^2$ between embedding vector $\mathbf{z}$ and centroid ${\bf c}_p$ of the normal class, are used to obtain the anomaly score.
First, we compute distance $d$ across the entire set of evaluation data, and then calculate the standardized distance $d^\prime$ (within the range of a maximum value one and a minimum value zero) across the entire data.
Finally, anomaly score $s$ is calculated using the following equation:
\begin{equation}
    \label{bce_ddcsad_ensemble}
    \begin{split}
        s = \alpha \times (1-p) + (1-\alpha) \times d^\prime,
    \end{split}
\end{equation}

where, ${\alpha}$ is a hyperparameter that determines the proportion of anomaly scores using posterior probability $p$.
\section{Experimental evaluation}
\subsection{Experimental conditions}
To evaluate the performance of the proposed method, we conducted an experiment using the DCASE~2020 Task~2~\cite{Koizumi_DCASE2020_01} data, which consists of two datasets, ToyADMOS~\cite{koizumi2019toyadmos} and MIMII~\cite{purohit2019mimii}.
From the ToyADMOS dataset, we used audio data for two types of machines, {\it ToyCar} and {\it ToyConveyor}, while MIMII provided audio data for four types of machines, {\it fan, pump, slider,} and {\it valve}, for a total of six machine types.
Each set of audio data for each type of machine consists of seven or eight different machines of that type, and ID information is provided to indicate exactly which machine the data belongs to.
For each machine (i.e., each ID), about 1,000 samples of normal sound are provided as training data, about 200 to 400 samples of normal and anomalous sounds from some of the machines are provided as validation data, and about 400 samples of normal and anomalous sounds from machines different from those included in the validation data are provided as evaluation data.
Each sample is about 10 seconds in duration, and includes the target machine’s operational and environmental sounds, with a sampling rate of 16 kHz on one recording channel.

\begin{table}[tb]
    \caption{$\alpha$ for BCE+DSAD and BCE+DDCSAD.}
    \label{table:ensemble_ratio}
    \centering
        \begin{tabular}{l|cccccc}
            \hline
            Method & fan & pump & slider & ToyCar & ToyConv. & valve \\ \hline
            BCE+DSAD & 0.1 & 0.2 & 0.0 & 0.0 & 0.0 & 0.0 \\
            BCE+DDCSAD & 0.1 & 1.0 & 1.0 & 0.1 & 0.0 & 1.0 \\
            \hline
        \end{tabular}
\end{table}

We compared the results when using each of the following five loss functions:
\begin{description}
    \item {\bf BCE}:
        A function which divides the feature space linearly.
        For the loss function, we used~\eqref{eq:bce}.
    \item {\bf DSAD}:
        The centroid of the normal data is defined to minimize within-class variance.
        For the loss function, we used~\eqref{deep_sad}.
        However, we did not use the pre-training model to initialize the AE's weight parameters, and we randomly initialized both the weight parameters and the parameters of the centroid of the normal data~$\bf c$~because we found that the random initialization tended to outperform the AE-based initialization.
    \item{\bf DDCSAD}:
        The centroids of both the normal and outlier data are defined to minimize within-class variance and maximize between-class variance.
        For the loss function, we used~\eqref{deep_dcsad}.
    \item {\bf BCE+DSAD}:
        We defined the centroid of the normal data to minimize within-class variance while creating a function that linearly divides the feature space.
        The loss function is derived by replacing $\mathcal{L}_\mathrm{DDCSAD}$ in~\eqref{eq:bce_ddcsad} with $\mathcal{L}_\mathrm{DSAD}$ in~\eqref{deep_sad}.
        In this case, as in the case of DSAD, we did not use pre-training to initialize the AE's weight parameters, and we randomly initialized both the weight parameters and centroid $\bf c$ of the normal data parameters.
    \item {\bf BCE+DDCSAD}:
        A function which divides the feature space linearly, while also defining the centroid for normal and outlier data to minimize within-class variance and maximize between-class variance.
        For the loss function, we used~\eqref{eq:bce_ddcsad}.
\end{description}
In order to accurately compare differences in performance related to the use of these various loss functions, we used the same pre-processing and network structure for all of the methods being compared, and only varied the loss function and presence of the full connection layer.
As a pre-processing step, we calculated each machine’s amplitude values, normalized them to have a mean of 0 and variance of 1, and then extracted 128-dimensional logarithmic Mel filter-banks, which were used as input features for the network using 1,024-sample windows and a hop-size of 512 samples.
The network structure consisted of a feature extractor with a convolutional layer that takes a series of acoustic feature as input, an aggregator that aggregates the acoustic feature and transforms them into fixed-length embedding vectors, and a full connection layer that performs binary classification using the embedding vectors.
As our feature extractor, we used the ResNet38 framework~\cite{he2016deep} proposed for pretrained audio neural networks (PANNs)~\cite{kong2020panns}.
Global average pooling, which averages in frequency and time information, was used as the aggregator.
We performed learning for each particular machine ID.
The normal data of the target ID of the target machine type was used as the normal data, and the normal data of other IDs of the target machine type, and the normal data of all the IDs of the other machines in the same data set, were used as outlier data~\cite{primus2020anomalous}.
The outlier data was used as pseudo-anomalous data in all of the methods compared.
We used different learning rates for each layer, 0.0001 for the convolutional layer and 0.001 for the full connection layer.
We used Adam~\cite{kingma2014adam} as the optimization method and multiplied the learning rate by 0.5 after every 1,000 iterations.
We set the total number of iterations to 4,000.
We set the value of $\lambda$ to 1.0 and the value of $\eta$ to 2.0.
During inference, we divided the number of frames in each sample’s acoustic feature series into ten sections, with overlap allowed so that the frame length was equal to 256.
We calculated anomaly scores for each segment of each series, and these scores were then averaged and used as the final anomaly score.
The values of $\alpha$ for inference were decided using the validation data, and are shown in Table~\ref{table:ensemble_ratio}.

\begin{table*}[tb]
    \centering
    \caption{Detection results AUC [$\%$] for all machines and loss functions ($95\%$ confidence interval~\cite{hanley1982meaning}).}
    \label{table:ex1_table}
        \begin{tabular}{c|ccccc}
            \hline
            Machine Type & BCE & DSAD & DDCSAD & BCE+DSAD & BCE+DDCSAD \\ \hline
            {\it fan} & $92.69 \pm 2.06$ & $82.33 \pm 3.00$ & $91.05 \pm 2.25$ & $92.96 \pm 2.02$ & $\bf 95.14 \pm 1.70$ \\
            {\it pump} & $\bf 94.39 \pm 1.88$ & $88.88 \pm 2.60$ & $91.71 \pm 2.27$ & $89.91 \pm 2.49$ & $92.14 \pm 2.21$ \\
            {\it slider} & $90.30 \pm 2.43$ & $93.46 \pm 2.01$ & $89.97 \pm 2.46$ & $93.38 \pm 2.03$ & $\bf 97.60 \pm 1.24$ \\
            {\it ToyCar} & $87.82 \pm 1.77$ & $84.10 \pm 2.00$ & $91.18 \pm 1.51$ & $83.72 \pm 2.03$ & $\bf 93.85 \pm 1.26$ \\
            {\it ToyConveyor} & $75.21 \pm 2.54$ & $68.51 \pm 2.74$ & $68.37 \pm 2.74$ & $64.13 \pm 2.82$ & $\bf 82.05 \pm 2.25$ \\
            {\it valve} & $89.92 \pm 2.45$ & $90.80 \pm 2.35$ & $88.62 \pm 2.59$ & $\bf 98.61 \pm 0.94$ & $96.08 \pm 1.57$ \\
            Machine Average & $88.39$ & $84.68$ & $86.82$ & $87.12$ & $\bf 92.81$ \\
            \hline
        \end{tabular}
\end{table*}

We used the area under the receiver operating characteristic (ROC) curve (AUC) as an evaluation metric, which is calculated as follows:
\begin{equation}
\footnotesize
    \label{auc}
    {\rm AUC} = \frac{1}{N_{-}N_{+}}\sum^{N_{-}}_{i=1}\sum^{N_{+}}_{i=j}{\cal H}\left({\cal A_{\theta}}\left({{\bf x}_j^{+}}\right)-{\cal A_{\theta}}\left({{\bf x}^{-}_i}\right)\right),
\end{equation}
where ${{\cal H}(a)}$ represents a binary function that returns 1 when $a>0$ and 0 when $a\leq0$, and where ${\cal A_{\theta}}\left({\bf x}\right)$ represents a function that returns an anomaly score when $\bf x$ is input.
${\{{\bf x}_i^{-}\}^{N^{-}}_{i=1}}$ and ${\{{\bf x}_j^{+}\}^{N^{+}}_{j=1}}$ represent normal and anomalous data, respectively, and are sorted to rank each sample’s anomaly scores in descending order.
$N_{-}$ and $N_{+}$ represent the number of normal and anomalous data samples, respectively.

\subsection{Performance evaluation when anomalous data is not used}
\label{ex1_results}
First, we investigated the performance of ASD when no anomalous data was used for training.
Our experimental results are shown in Table~\ref{table:ex1_table}.
The averaged results (Machine Average) in Table~\ref{table:ex1_table} show that the DDCSAD loss function outperformed the DSAD loss function.
This result suggests that it is important to consider not only the centroid of normal data but also the centroid of outlier data in order to increase between-class variance, which is achieved by making the centroids updatable.
Furthermore, the improvement in performance when using BCE+DDCSAD over that of using either DDCSAD or BCE alone confirms the effectiveness of multi-task learning.

\subsection{Relationship between amount of anomalous training data and performance}
\label{ex2}
We added a small amount of anomalous data to the training data to investigate how this affected performance.
For each machine ID, 64 samples were randomly selected from the validation and evaluation data’s anomalous data, and moved to the training data.
We then increased the number of anomalous samples used for training to $[1,2,4,...,64]$ for each method and observed the performance change.
The pre-processing, learning and inference procedures were identical to those described in \ref{ex1_results}, but we made the following two changes in the event that new anomalous data became available during the operation of the ASD system:
\begin{enumerate}
    \item The ratio of normal data, outlier data and anomalous data was set to 32:31:1 so that there was always one anomalous data sample representing the outlier class in each mini-batch.
    \item We used the model trained without anomalous data as the initial value, and then halved the total number of iterations to 2,000.
\end{enumerate}
\begin{figure*}[t!]
    \captionsetup[subfigure]{position=b}
    \centering
    \begin{subfigure}[b]{0.3\textwidth}
        \includegraphics[clip, width=5.5cm]{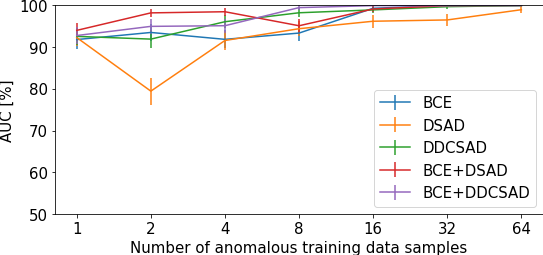}
        \caption{{\it fan}}
    \end{subfigure}
    \begin{subfigure}[b]{0.3\textwidth}
        \includegraphics[clip, width=5.5cm]{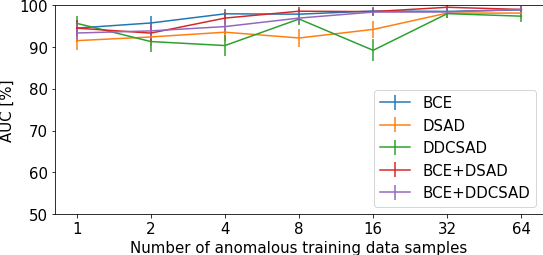}
        \caption{{\it pump}}
    \end{subfigure}
    \begin{subfigure}[b]{0.3\textwidth}
        \includegraphics[clip, width=5.5cm]{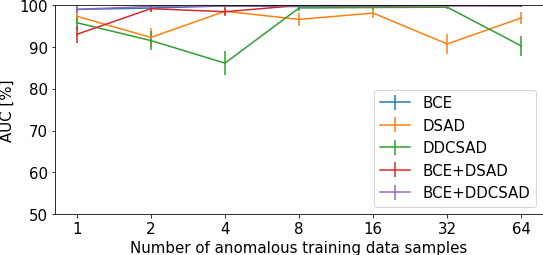}
        \caption{{\it slider}}
    \end{subfigure}
    \begin{subfigure}[b]{0.3\textwidth}
        \includegraphics[clip, width=5.5cm]{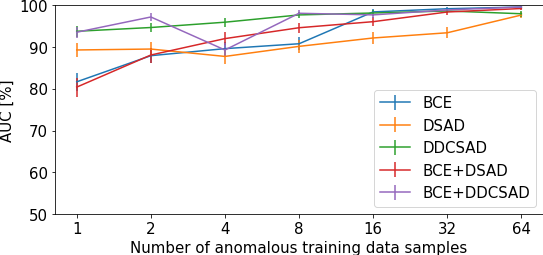}
        \caption{{\it ToyCar}}
    \end{subfigure}
    \begin{subfigure}[b]{0.3\textwidth}
        \includegraphics[clip, width=5.5cm]{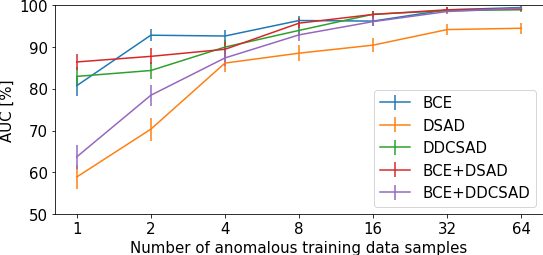}
        \caption{{\it ToyConveyor}}
    \end{subfigure}
    \begin{subfigure}[b]{0.3\textwidth}
        \includegraphics[clip, width=5.5cm]{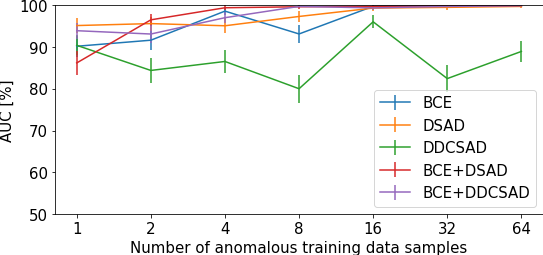}
        \caption{{\it valve}}
    \end{subfigure}
    \caption{Change in AUC [$\%$] for each machine and loss function when adding anomalous data ($95\%$ confidence interval~\cite{hanley1982meaning}).}
    \label{fig:all_auc_move-eval-checkpoint-6000steps}
\end{figure*}

Experimental results when using the evaluation data are shown in Fig.~\ref{fig:all_auc_move-eval-checkpoint-6000steps}.
We can see from these results that each method’s performance improves as more anomalous training data is added.
In other words, we can improve the performance of all of these methods by simply adding anomalous data to the training data without changing the systems’ structure.
Furthermore, by comparing methods with and without BCE, we can also confirm that BCE-based methods receive a more dramatic boost in performance by utilizing anomalous data during training.
This suggests that the proposed multi-task learning makes more effective use of small amounts of anomalous training data.
\section{Conclusion}

In this paper, we have proposed a multi-task learning method for detecting anomalous sounds which uses a binary classification model based on outlier data, and a loss function based on metric learning.
We also proposed DDCSAD, a loss function that considers the class centroids of both normal and anomalous data in the feature space, and demonstrated its effectiveness through experimental evaluation.
The relationship between the amount of anomalous data used during training and detection performance was also investigated.
Our experimental results showed that the more anomalous data added to the outlier class during training, the better detection performance becomes, especially when using a binary classification model.
In future work, we will develop a metric for selecting outliers to use for training from large data sets to further improve performance.


\section*{Acknowledgment}
This paper was partly supported by a project, JPNP20006, commissioned by NEDO.

\bibliographystyle{IEEEtran}
\bibliography{ref}
\end{document}